\documentclass[journal]{IEEEtran}

\ifCLASSINFOpdf
  \usepackage[pdftex]{graphicx}
  \DeclareGraphicsExtensions{.pdf,.jpeg,.png,.jpg}
\else
  \usepackage[dvips]{graphicx}
  \DeclareGraphicsExtensions{.eps}
\fi
\usepackage[cmex10]{amsmath}

\usepackage{mathrsfs}
\usepackage{xy}
\xyoption{all}

\usepackage{amsthm}
\usepackage{amssymb}
\theoremstyle{plain}

\newtheorem{thm}{Theorem}

\theoremstyle{definition}

\def\argmin{\mathop{argmin}}

\newcommand{\Design}{{\mathcal D}}        
\newcommand{\Configuration}{{\mathcal C}} 
\newcommand{\Excitation}{{\mathcal E}}    
\newcommand{\Full}{{\mathcal F}}          
\newcommand{\Pattern}{{\mathcal P}}       
\newcommand{\Measurement}{{\mathcal M}}

\begin{document}
%

\title{Knowledge-based antenna pattern extrapolation}
%
%
%

\author{Michael~Robinson$^{(1)}$,~\IEEEmembership{Member,~IEEE}
\thanks{(1) SRC, inc., 6225 Running Ridge Road, N. Syracuse, NY 13212 USA and University of Pennsylvania, 209 S. 33rd Street, Philadelphia, PA, 19104, USA e-mail: mrobinson@srcinc.com,
robim@math.upenn.edu}}
\markboth{IEEE Transactions on Antennas and Propagation}{Knowledge-based antenna pattern extrapolation}
%


\maketitle

\begin{abstract}
We describe a theoretically-motivated algorithm for extrapolation of antenna radiation patterns from a small number of measurements.  This algorithm exploits constraints on the antenna's underlying design to avoid ambiguities, but is sufficiently general to address many different antenna types.  A theoretical basis for the robustness of this algorithm is developed, and its performance is verified in simulation using a number of popular antenna designs.
\end{abstract}

\begin{IEEEkeywords}
antenna pattern; antenna extrapolation; design space; Whitney embedding 
\end{IEEEkeywords}

%

\section{Introduction}

\IEEEPARstart{M}{easurement} of antenna radiation patterns requires extensive sampling of transmitted or received signals at many angular locations.  Scanning a known antenna over the antenna under test is the usual approach, but this can be time-consuming.  We propose a theoretical basis and a practical algorithm for using a small number of measurements to estimate the {\it unseen and unmeasured} portions of the antenna's radiation pattern, by exploiting knowledge about the antenna's design.  The method is robust enough to detect when the antenna's design model is incorrect, yet general enough to treat many different popular classes of antennas without essential changes. 

Although the inverse source problem is unsolvable in full generality, it is solvable if the configurations of radiating and passive structures are known to be constrained.  In particular, a solution of the inverse source problem would infer a volumetric current distribution from extensive pattern measurements.  Both current distributions and radiation patterns are described by infinite dimensional vector spaces: this leads to nonuniqueness.  The {\it design spaces} we propose have the distinct advantage that they are finite dimensional descriptions of possible antennas, and therefore constrain the inverse source problem to a {\it highly overconstrained} problem in its most ideal incarnation.  Since radiation pattern of an antenna varies smoothly as its design is adjusted, the well-known Whitney embedding theorem\footnote{for a detailed treatment of the necessary differential topological background, see Lee \cite{Lee_2003}} applies, and indicates that the inverse source problem in this restricted setting is solvable.  Further, it yields bounds on the number of samples one must take from the radiation pattern in order to solve for the antenna's design in general.  

Inspired by this general theoretical result, we describe a general algorithm that exploits the measurements taken of an antenna and its associated design pattern in order to estimate the pattern at arbitrary directions.  In doing so, we carefully treat the problem of symmetries that arise both in the antenna's configuration and in the structure of the measurement sample locations.  If unchecked, symmetric sampling methods can result in ambiguities in the radiation pattern estimates, but they are generally easy to avoid.

\section{Historical development}

Recovery of an antenna's excitation from a limited number of pattern measurements, and subsequent prediction of additional portions of the pattern has been explored by many other authors.  In its usual incarnation (for instance, as described by \cite{Gregson_2006}), the antenna is scanned in its near field.  After careful calibration for mutual coupling effects, near field scanners obtain a very detailed model of the antenna's current distribution.  From this current distribution, accurate models of the resulting antenna pattern can be predicted.  A disadvantage of near-field scanning is that sophisticated, expensive equipment and careful calibration are required.  (It should be noted that if the scanner is designed to fit exactly one specific antenna, some equipment complexity can be removed \cite{SRC_scanner}.)  In contrast, our approach is general enough to handle many different antenna types essentially without modification, can use simple equipment (a field strength measurement meter suffices), and has minimal calibration requirements.  Since the measurements can be taken in the far field, mutual coupling is essentially eliminated.  (Our methodology also works in the near field, though the advantages are less clear in that case.) 

Another article with a similar processing chain to ours is \cite{CanoFacila_2011}.  In this article, the authors use an explicit far-to-near field transformation to filter far-field measurements.  In particular, since they are working exclusively with planar antennas, any excitations found to be off of the plane consitute error, and are removed.  In essence, they manipulate the excitation structure of the antenna under the assumption of a known configuration.  Their algorithm is therefore linear.  Our work is an extension of this idea, and allows antenna configuration to vary, resulting in the need for {\it nonlinear} optimization.

More theoretically, the successful extrapolation of antenna patterns relies on the uniqueness of antenna current distributions given a subsampling of the pattern.  This was an open problem for some time, until solved in the {\it negative} by Gbur in his dissertation \cite{Gbur_2001}.  The particular difficulty is that there is too much freedom in specifying the current distribution.  There exist certain distinct current distributions that result in the same radiation pattern.  These appear to be highly symmetric distributions.  They are also nongeneric: patterns that are small perturbations of the offending ones uniquely specify current distributions.

We argue that the problem becomes solvable if the currents are appropriately constrained; we show that this follows using the techniques of differential topology.  Because of the generality of the argument we follow, it is possible to choose many different parametrizations of the current distributions.  Because it is most congenial to the design of antennas, we divide the parameters into two families: those that describe the relative magnitude and phase of signals applied to the antenna's ports ({\it excitations} in what follows), and those that describe the relative positions and sizes of reflecting and radiating elements (the antenna's {\it configuration}).  

\subsection{Contrasting our approach with previous work}

Unlike the research previously discussed, our approach is substantially more general.  Since it is borne of a {\it topological} treatment of the problem, our approach is generically insensitive to the type of antenna and associated measurement campaign.  Algorithms based on this topological theory of radiation patterns are therefore {\it flexible enough to handle any kind of antenna}, requiring the authoring of only a pattern computation model and a representation of the antenna design space.  Our algorithms demonstrate good performance on a wide variety of interesting antennas, rather than being effective only on a particular kind of antenna.  Although we frame these examples from the point of view of far-field, scalar measurements, this is by no means necessary.  Indeed, our methods include polarization-sensitive measurements and near-field scanning as special cases.

\section{General theory}
In this section, we develop the general theory that permits extrapolation of antennna patterns.  We begin by giving precise descriptions of antenna design spaces and antenna patterns, then we apply the Whitney embedding theorem to this setting, and finally we describe a practical algorithm derived from this theoretical grounding. 

Our approach relies on the fact that when enough pattern measurements are taken, the radiating structure of the antenna is completely constrained.  Given this radiating structure, we can extrapolate the directivity at any location.  Of course, a crucial point is that it takes finitely many parameter to specify a particular antenna in a class of possible designs.

\subsection{Design and measurement spaces}
\label{sec:design}

A {\it design space} is a product $\Configuration\times\Excitation$ of a finite dimensional manifold $\Configuration$ (possibly with boundary) called the {\it configuration} and a vector space of complex $N$-tuples $\Excitation=\mathbb{C}^N$, the {\it excitation}.  We interpret the configuration as defining the parameters of an antenna's design: perhaps the element positions, locations of reflectors, type of grounding, etc.  The excitation consists of the relative magnitudes and phases being applied to each active element (or feed port) of the antenna.  

\begin{figure}
\begin{center}
\includegraphics[width=2in]{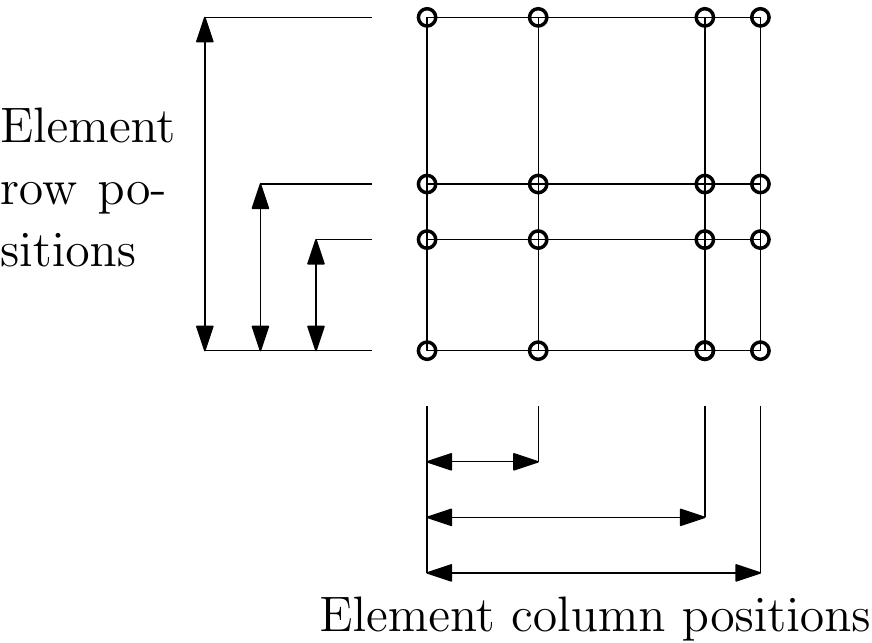}
\end{center}
\caption{Schematic of configuration space for rectangular phased arrays}
\label{rect_conf_fig}
\end{figure}

\begin{figure}
\begin{center}
\includegraphics[width=2in]{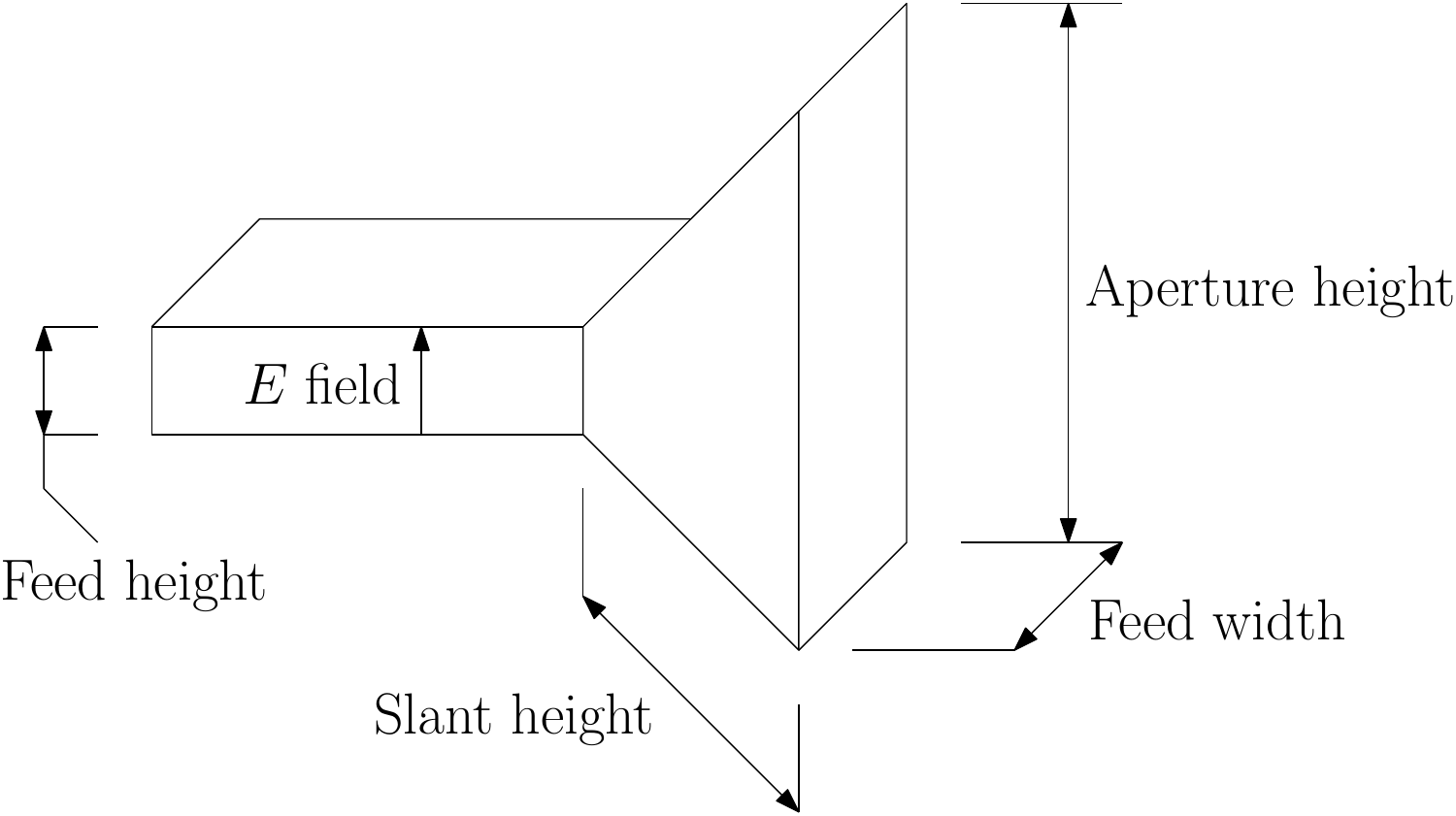}
\end{center}
\caption{Schematic of configuration space for E-plane horn}
\label{horn_conf_fig}
\end{figure}

\begin{figure}
\begin{center}
\includegraphics[width=2in]{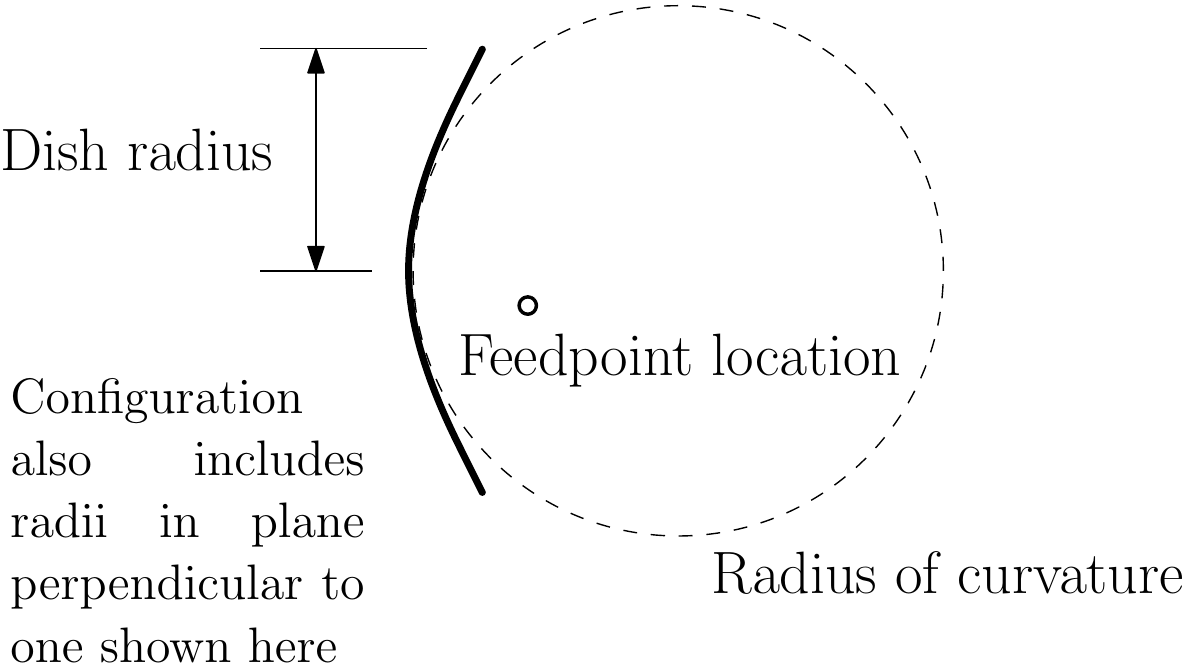}
\end{center}
\caption{Schematic of configuration space for isotropic-fed parabolic dish}
\label{dish_conf_fig}
\end{figure}

Here are some typical examples of antenna designs and their associated design spaces: 
\begin{enumerate}
\item General phased array with $N$ elements (location of each, overlapping elements are permitted):  $\Configuration=\mathbb{R}^{3N}$, and $\Excitation=\mathbb{C}^N$.
\item Rectangular array (see Figure \ref{rect_conf_fig}) with $M$ rows and $N$ columns: $\Configuration=\mathbb{R}^{M+N-2}$, and $\Excitation=\mathbb{C}^{MN}$.  (Note that the number of row spacings is $M-1$ and the number of column spacings is $N-1$.)
\item Horn (see Figure \ref{horn_conf_fig}): $\Configuration=\mathbb{R}^3$ and $\Excitation=\mathbb{C}$.
\item Dish (see Figure \ref{dish_conf_fig} for a picture of a planar slice of the antenna; the full configuration includes the perpendicular radius and curvature, as well as orientation of feed).  Assuming a fixed, known feed with an elliptical dish cut from a paraboloid: $\Configuration=\mathbb{R}^2\times\mathbb{R}^2\times\mathbb{R}^3\times SO(3)$ and $\Excitation=\mathbb{C}$.  ($SO(3)$ is the group of orthogonal 3x3 matrices with unit determinant, and also the group of rotations in 3 dimensions.) 
\end{enumerate}

The primary focus of this article is an antenna's {\it pattern}, which we axiomize as a smooth map $\Full$ from the design space $\Design$ to a space of {\it measurements} $\Measurement$, a Banach space of $V$-valued functions on the unit sphere.  We assume $V$ is a $\mathbb{C}$-vector space, describing the polarization response in the far field.  We assume that $\Full$ is linear on the $\Excitation$ factor, and may be (usually is) nonlinear on $\Configuration$.  Generally speaking, $V$ can be less informative depending on the kind of pattern under study: $V=\mathbb{R}$ for a magnitude-only pattern, $V=\mathbb{C}$ for a pattern in magnitude and phase, $V=\mathbb{C}^3$ for a fully-polarized response.  

Of course, $\Full$ is a theoretical construct only; in practice one takes measurements of the pattern at finitely many directions.  Each one of these {\it sample points} is an element of $\{\phi_i\}_{i=1}^K \subset S^2$.  In this case, the full pattern $\Full$ restricts to the {\it (sampled) antenna pattern} $\Pattern:\Design\to V^K$.

One should be aware that our model does not explicitly incorporate any physical model of propagation, mutual coupling, or excitation.  The approach discussed here is {\it topological} and thereby insensitive to the physics beyond what we assumed above.  As will be clear in the next section, the critical resource for extrapolating patterns is the dimension of the number of sample points $K$ versus the dimension of $\Design$.

\subsection{Embedding of the design space}

The obstruction to determining an antenna's design from measurements is lack of injectivity of the pattern $\Pattern$.  However, it is a general fact that $\Pattern$ is injective when there are enough sample points.

\begin{thm}
\label{thm:uniqueness}
Suppose that $2\dim \Design < K \dim V$.  For generic choice of $\Pattern \in C^\infty(\Design; V^K)$, there exists a smooth \emph{extrapolation map} $\Full \circ (\Pattern)^{-1}:\Pattern(\Design) \to \Measurement$ from a submanifold of $V^K$ to $\Measurement$.  
\end{thm}

In particular, this map takes a sampled pattern in the image of $\Pattern$ to the pattern extended to all possible sample point locations.  This map is well defined and smooth for generic choices of $\Pattern$.

\begin{IEEEproof}
First, observe that the pattern $\Pattern:\Design\to V^K$ is a topological embedding (hence injective) under a generic perturbation in $C^\infty(\Design;V^K)$ when $2\dim \Design < K \dim V$.  This follows immediately from the Whitney embedding theorem as given in \cite{MROD} for manifolds with boundary, since $\Pattern$ is a smooth map.  One should notice that although we assumed smoothness of $\Pattern$ (dependence of the pattern on the configuration), in fact we have obtained smoothness of the inverse.

It then follows immediately that $\Full$ is generically an embedding, since $\Measurement$ is an infinite dimensional vector space.
\end{IEEEproof}

The reader is cautioned that genericity of $\Pattern$ is not a completely innocuous condition.  Precisely, it indicates that in the vast majority of situations, patterns will uniquely determine antenna configuration and excitation.  However, this implies that in the situations where $\Pattern$ is injective, there is some measure of symmetry-breaking.  

A simple example of a choice of $\Pattern$ for which the conclusion of Theorem \ref{thm:uniqueness} does {\it not} hold arises from sampling azimuth patterns only.  Specifically, let us consider the case of a phased array composed of $N$ isotropic radiators located at $\{x_m\}_{m=1}^N$ with excitations $\{a_m\}_{m=1}^N$ respectively.  Consider the pattern map arising from taking sample points at $s_k=(R\cos k\theta,R\sin k\theta, 0)$ (where $R, \theta$ are fixed), namely
\begin{equation*}
\Pattern(x_1,...,x_N,a_1,...,a_N)_k=\sum_{m=1}^N a_m e^{i\omega \|x_m-s_k\|/c}.
\end{equation*}
There is an inherent ambiguity whether a given $x_m$ is above and below the $xy$-plane.  On the other hand, many small vertical perturbations of the sample points will induce a small perturbation of the resulting pattern maps (since $\exp$ is smooth), yet will {\it break the symmetry} and thereby ensure the existence of a extrapolation map.  (Proving injectivity of $\Pattern$ in this case is an elementary, if tedious, exercise for the reader.) 

Because symmetric (or nearly symmetric) sampling is common in practice, the existence of a extrapolation map could be an inherently delicate phenomenon.  In order to validate its robustness in practice, extensive simulation is discussed in Section \ref{sec:sim} with realistic antenna configurations and various sampling patterns.

\subsection{Algorithm for extrapolating the pattern}

In this section, we describe an algorithm\footnote{A patent is pending on the algorithm} for computing and approximation to the inversion $\Pattern^{-1}:\Pattern(\Design) \subset V^K\to \Design$.  In this way, we determine the configuration and excitation of the antenna from its pattern.  Since $\Design$ may have a large dimension, and $\Pattern$ may have a complicated form, it is desirable to devise a method that is flexible, yet exploits structure in $\Pattern$.  To this end, we use the product decomposition $\Design=\Configuration\times\Excitation$, and the fact that $\Pattern$ depends linearly on the $\Excitation$ factor.  This has two major benefits: (1) constrained least-squares can be used to robustly recover the element of $\Excitation$ given a known element of $\Configuration$, and (2) the nonlinearity of $\Pattern$ is confined to a much lower dimensional factor, namely $\Configuration$.  We can therefore use a gradient-like search algorithm to traverse $\Configuration$ in a robust fashion.  

Our algorithm takes as input 
\begin{itemize}
\item A parametric representation of the antenna design space $\Design$, specifically written as the product $\Configuration\times\Excitation$, 
\item The locations of each sample point,
\item Signal measurements at each sample point, 
\item A sample point for which an estimated measurement is desired, and
\item An implementation of the pattern predictor $\Full$.
\end{itemize}
As output, it generates an estimate of the measurement at desired sample point.

The algorithm proceeds recursively, starting with $(x_0,a_0)\in\Configuration\times\Excitation$ and an associated set of measurements $p\in\Measurement$, and producing a sequence $(x_n,a_n)$ which we compute until some convergence criterion is met.  Each iteration is computed from the iteration immediately before it, in the following way.  

\begin{enumerate}
\item A dense rectangular grid with a fixed number of elements of $\Configuration$ is computed, called $\{x^j_n\}_j$, centered on $x_n$.  This grid's orientation depends on the particular parametrization of $\Configuration$ that was chosen for the antenna under test.  The spacing between adjacent $x^j_n$ elements is chosen to be a decreasing function of $n$.
\item For each element of $\{x^j_n\}$, the best (in the least squares sense, constrained to unit magnitude) excitation $a^j_n$ associated to the {\it linear} problem $\Full(x^j_n,a^j_n)=p$ is computed.
\item The next iterate $(x_{n+1},a_{n+1})$ is defined to be $\argmin_{(x^j_n,a^j_n)} \|\Full(x^j_n,a^j_n)-p\|$.
\end{enumerate}

\section{Simulation}
\label{sec:sim}

We have validated performance of our algorithm using appropriately parametrized design spaces $\Design$, as described in the previous section.  These results confirm both the theoretical and practical value of our method, by showing that the pattern map $\Pattern$ is injective in these cases, and can be used to robustly extrapolate to an approximation to $\Full$.

\subsection{Sampling patterns}
\begin{figure}
\begin{center}
\includegraphics[width=1.5in]{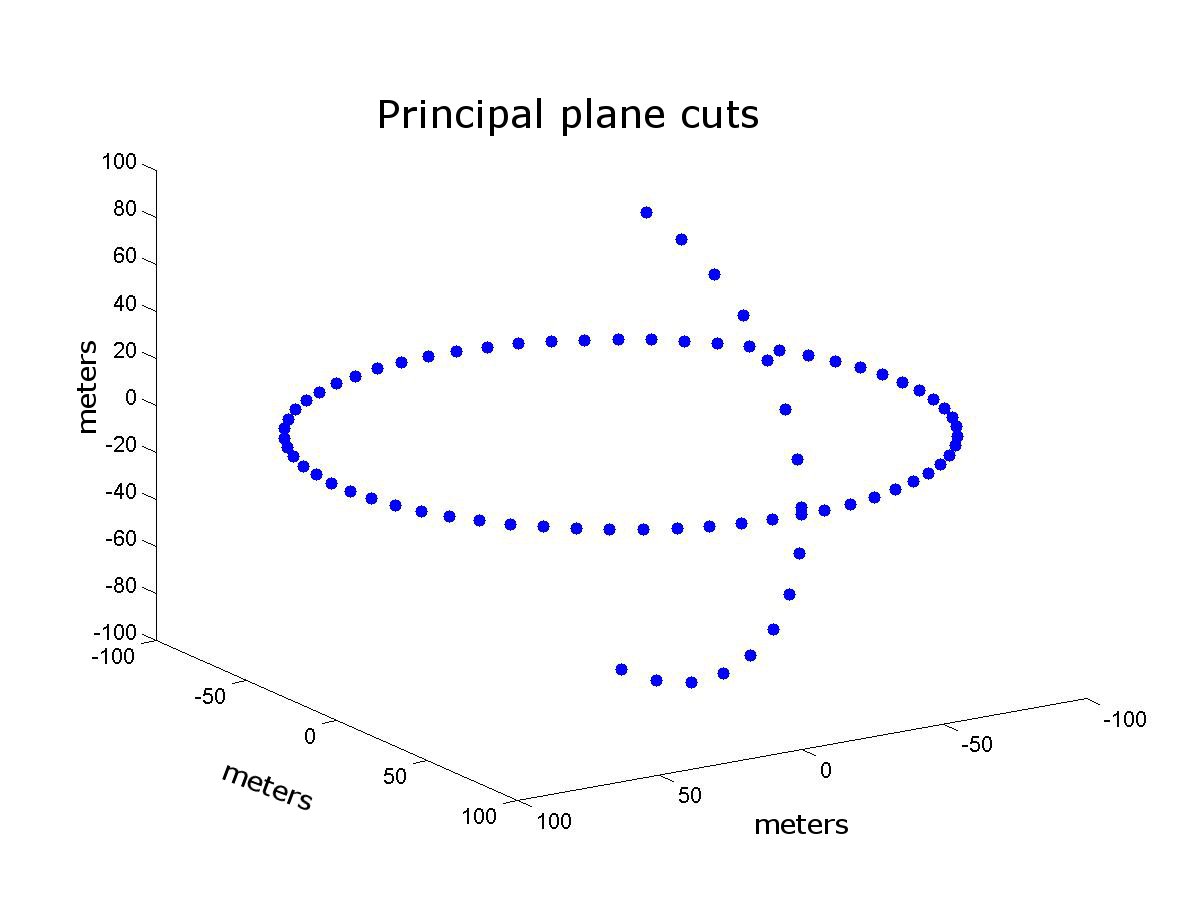}
\includegraphics[width=1.5in]{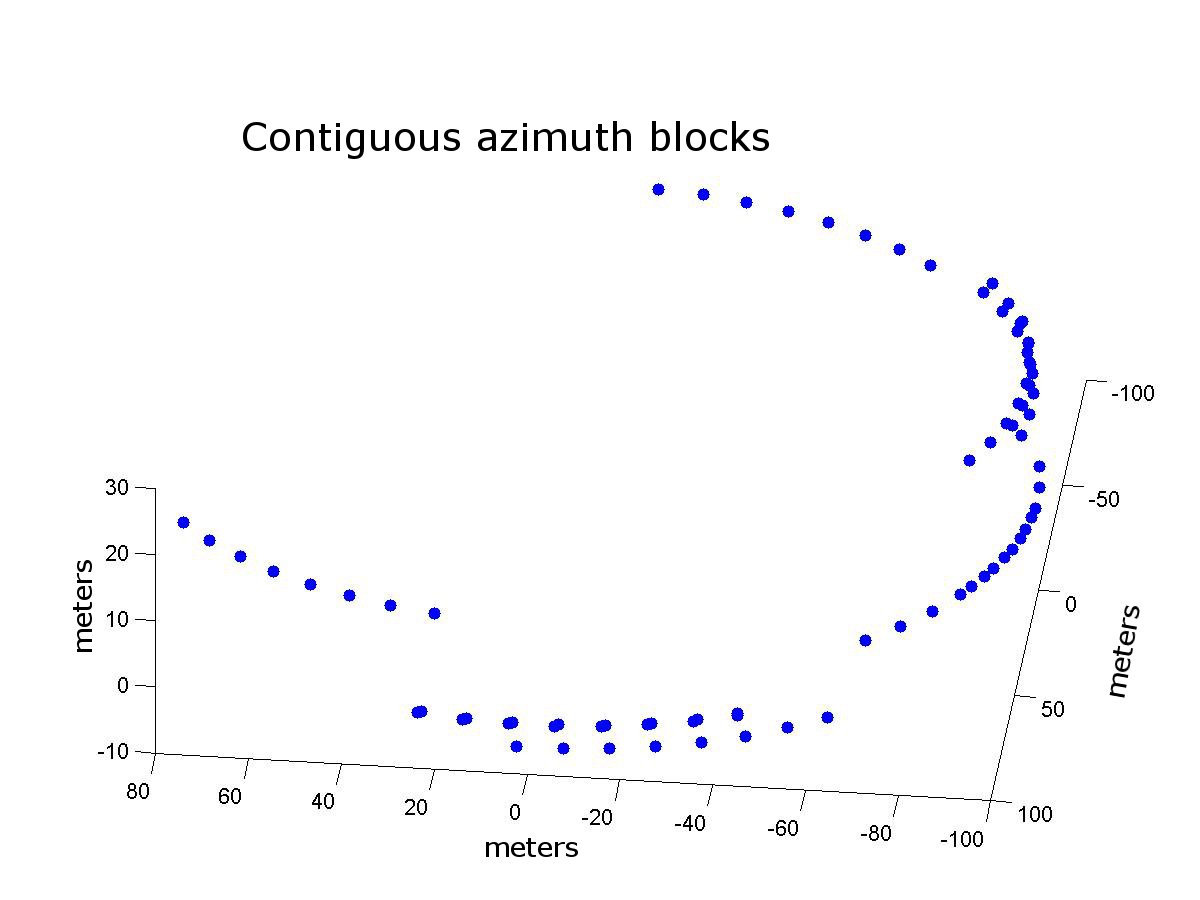}
\includegraphics[width=1.5in]{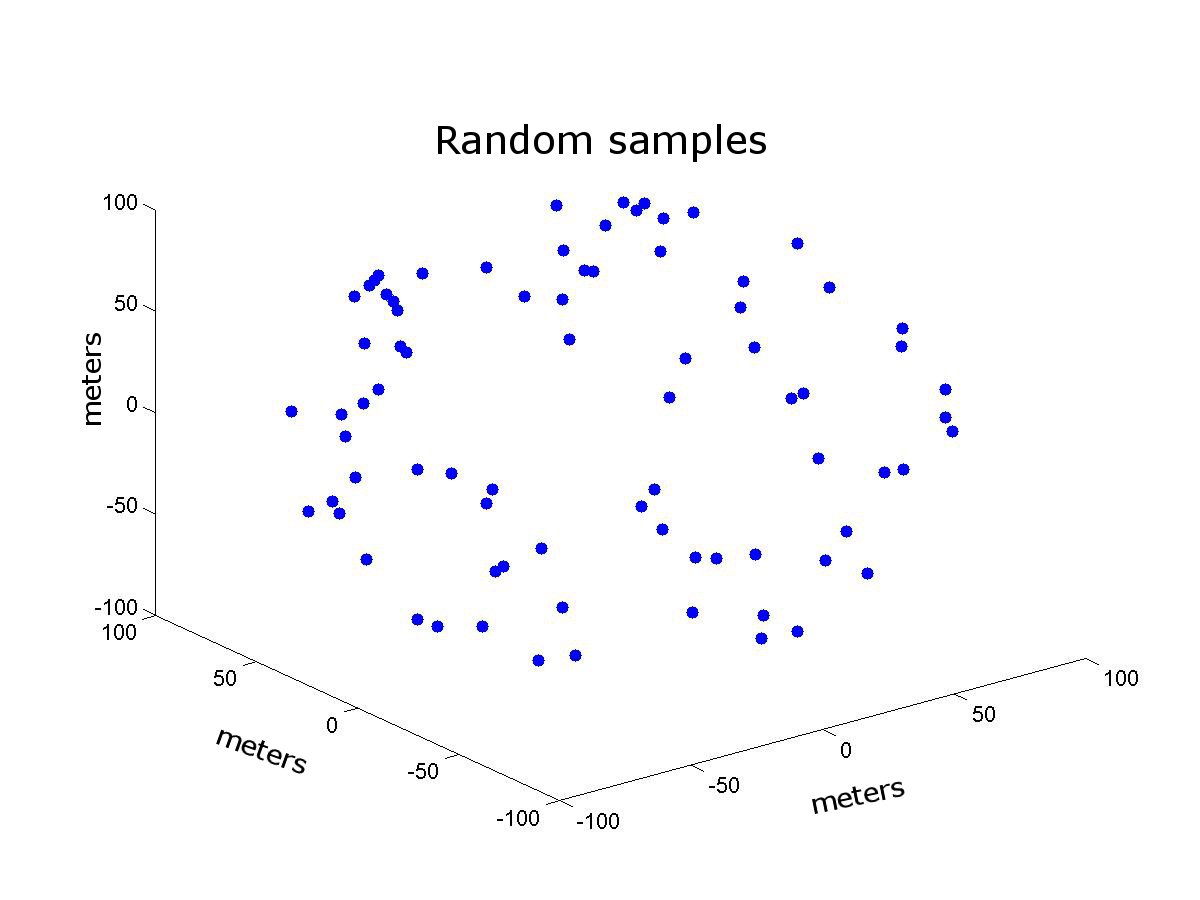}
\end{center}
\caption{Spatial layout of sampling patterns used in our study, clockwise from top left: compete azimuth and elevation samples, blocks of contiguous azimuth samples, random samples.  Axes are in meters.  Not shown: azimuth samples only }
\label{sampling_fig}
\end{figure}

In order to determine sensitivity of our algorithm to the choice of sample points, we tried several kinds of sample point layouts.  Specifically, we examined (see Figure \ref{sampling_fig})
\begin{enumerate}
\item azimuth pattern samples only, 
\item blocks of contiguous portions of azimuth samples, at random small elevation angles, 
\item complete azimuth and elevation patterns together, 
\item and random sampling.
\end{enumerate}
It's worth noting that although azimuth and elevation patterns are considered standard procedure due to the mechanical simplicity of collecting them, they can be suboptimal for our techniques.  Specifically, if the azimuth- or elevation-plane is a plane of symmetry for the antenna, ambiguities in $\Full$ may arise.  Therefore, to balance between measurement complexity and the need to break sampling symmetry, we also considered contiguous blocks of azimuth samples at random elevation angles.  We compared the performance of our approach using a large number of typical antennas to uncover the relationship between sampling patterns and the resulting ambiguities. 

\subsection{Performance metrics}

Our experimental procedure aims to validate the performance of the extrapolation algorithm by comparing its output to simulated truth.  We did this by generating a densely sampled pattern at many azimuth and elevation angles from a known configuration.  From this densely sampled pattern, the subset of samples to be used as input to the algorithm was selected.  The algorithm was then applied to estimate the pattern at each of the original sample points.  Although directly computing the differences between the estimated and truth values gives a measure of performance, the {\it total error} is not very useful in practice, since the truth pattern is unavailable. 

A more predictive performance measure exploits the fact that the estimated pattern differs from the suppled (undersampled) pattern.  This pattern {\it residual error} (between output of the algorithm and the input pattern) should ideally be zero, but due to discretization and numerical errors does not vanish.  By comparing the residual error and the total error, it becomes possible to predict the performance of the algorithm knowing only the residual error.  Indeed, in our example cases, reliable estimates can be obtained.  For instance, consider Figures \ref{rect_extrap_fig}, \ref{horn_extrap_fig}, and \ref{dish_extrap_fig}.  As the residual (the difference between supplied pattern and the predicted pattern at the sample points) decreases (moving leftward in Figures \ref{rect_extrap_fig}, \ref{horn_extrap_fig}, and \ref{dish_extrap_fig}), it is clear that the total error in the pattern generally decreases.  

It should be noted that the antenna pattern measurement space $V^K$ can be a vector space over $\mathbb{C}$ (magnitude and phase) or $\mathbb{R}$ (magnitude).  In the latter case, it is theoretically immaterial whether the magnitude is given in absolute units or in decibels.

\begin{figure}
\begin{center}
\includegraphics[width=3in]{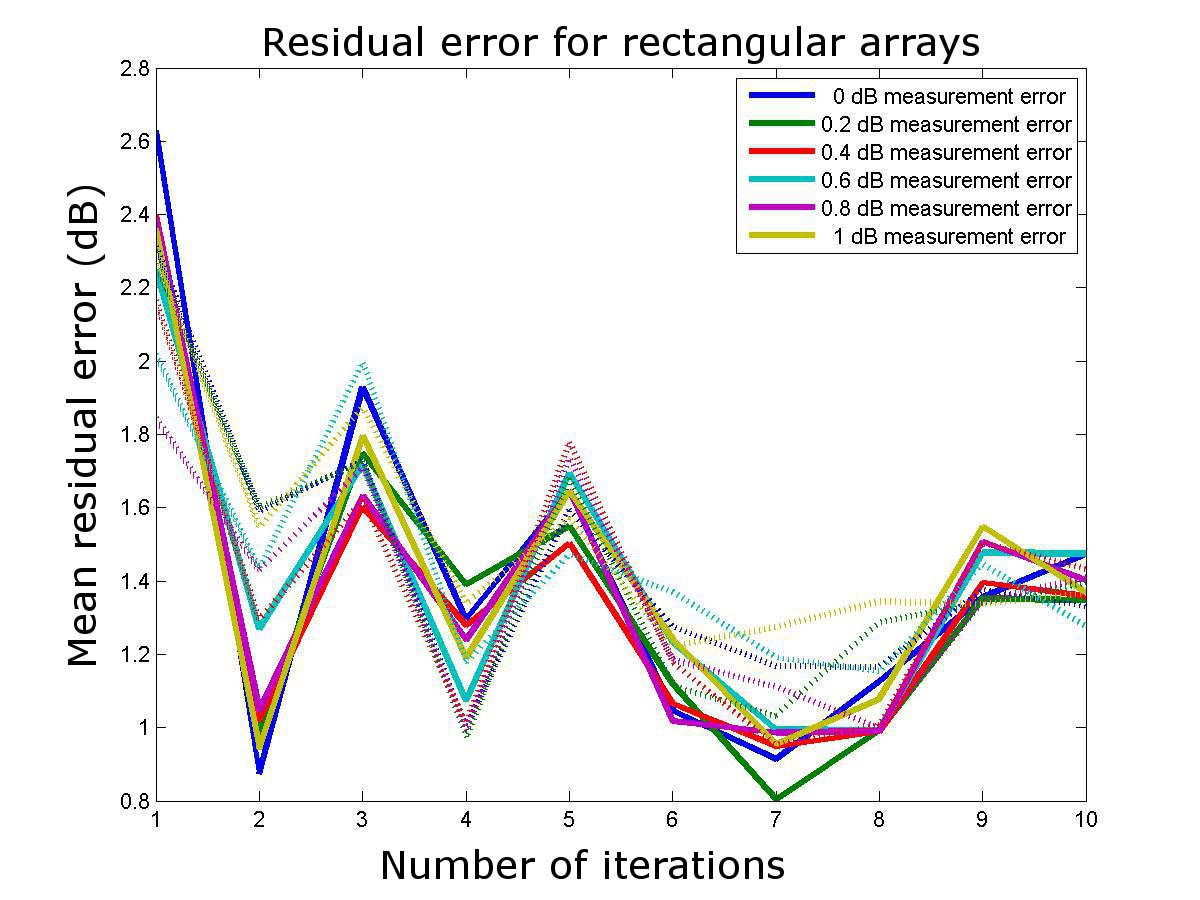}
\end{center}
\caption{A typical evolution of residuals over solver iterations for a 3x3 rectangular phased array.  Varying amounts of measurement error was added to the sampled patterns to test robustness of convergence.}
\label{iterations_fig}
\end{figure}

Additionally, one can examine the residual errors and their evolution over the course of the iterations of the algorithm.  In Figure \ref{iterations_fig}, several different amounts of additive Gaussian white noise was added to the pattern measurements, and the resulting degradation of residual error was recorded.  Since the residual error decreases over the course of the algorithm's execution without much dependence on measurement error, we deem the algorithm to be robust to these errors.

Residual error evolution can also be used to identify inconsistences in the choice of design space and thereby make refinements in the model.  This permits substantially greater power than is indicated by the theory.  For instance, if a phased array is correctly postulated for the unknown antenna's configuration, but the wrong number of elements is specified, it will be difficult to reduce the residual to zero.  Supplying the correct number of elements permits the algorithm to converge.  

\section{Results}

In this section, we discuss a number of special cases of Theorem \ref{thm:uniqueness}.  We indicate what the configuration space is for the particular antenna in question, as well as parametrizations we have found convenient for use in our algorithm.

To keep the exposition simpler, we treat idealized models of antennas, ignoring polarization, ground effects, support structures, and dielectrics.  The reader may supply more realistic models as appropriate.

\subsection{Rectangular phased arrays}

As noted in Section \ref{sec:design}, rectangular phased arrays have a design space given by $\Configuration=\mathbb{R}^{M+N-2}$, and $\Excitation=\mathbb{C}^{NM}$, where there are $M$ rows and $N$ columns of elements.  The components of $\Configuration$ specify the element spacing, as shown in Figure \ref{rect_conf_fig}.  In this case, the full pattern map is straightforward to construct, namely
\begin{equation*}
\Full(x,y,a)(\phi,\theta)=\sum_{j=1}^M\sum_{k=1}^N a_{jk} e^{i \omega (y_j \sin\theta + x_k \sin\phi )/c},
\end{equation*}
where $x_k$ represents the column locations ($x_1=0$ is fixed), $y_j$
represents the row locations ($y_1=0$ is fixed), and the antenna
elements lie in the $xy$-plane.

It is known that the full pattern of a given rectangular phased array is uniquely determined by its excitations.  This follows from the fact that the near-to-far transform (taking $\Excitation\to L^2(S^2,\mathbb{C})$) is a discrete Fourier transform under an appropriate change of variables, and the Fourier transform is invertible.  (See \cite{Sarkar_1999} for a standard treatment.)  Our Theorem \ref{thm:uniqueness} extends this standard result to treat subsampled patterns, and our algorithm can treat this situation.  Generically, one would need to take $K$ measurements, where $K>2+2MN$.

\begin{figure}
\begin{center}
\includegraphics[width=3in]{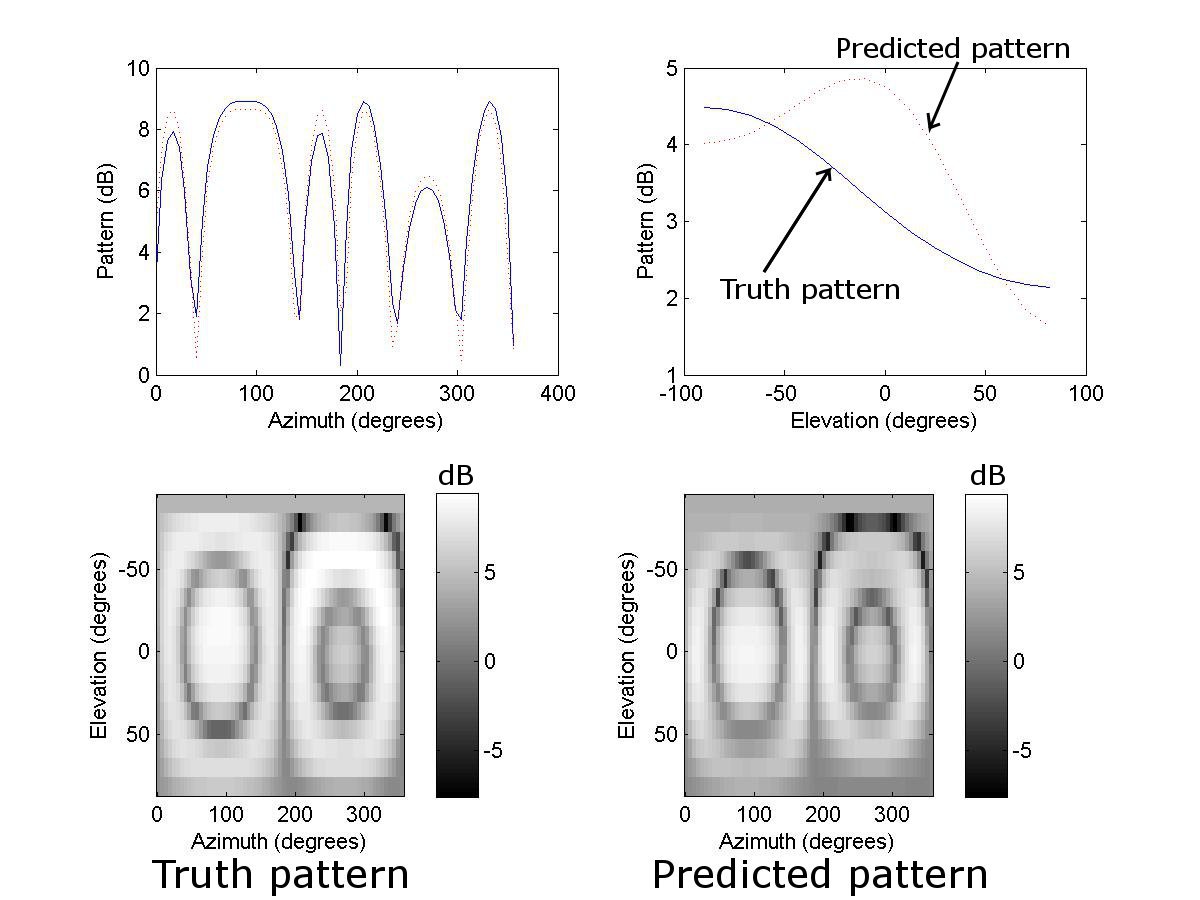}
\end{center}
\caption{Typical rectangular array pattern.  Top frames show azimuth and elevation patterns respectively; dotted lines represent the closest match found by our algorithm.  Bottom left frame shows truth pattern, bottom right is the prediction from our algorithm.}
\label{rect_eg_fig}
\end{figure}

\begin{figure}
\begin{center}
\includegraphics[width=3in]{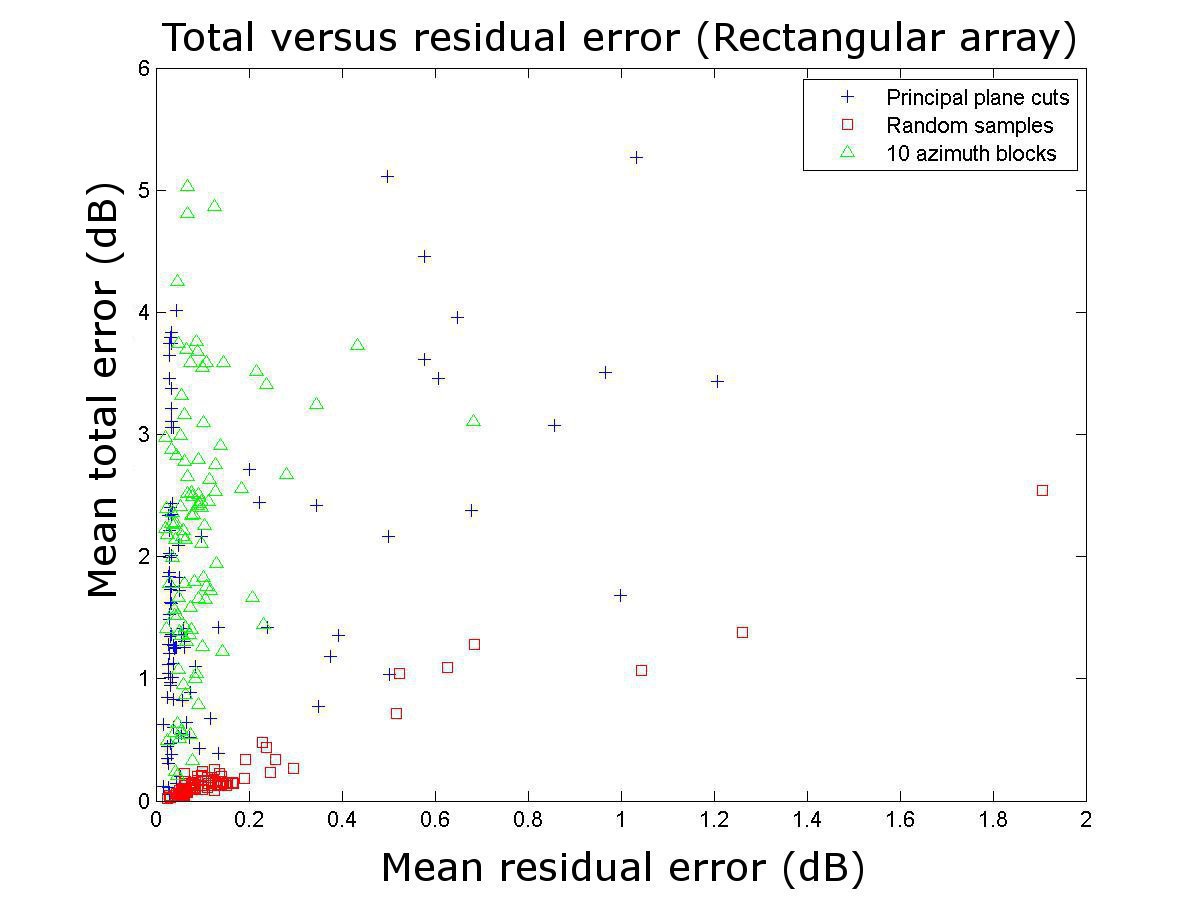}
\end{center}
\caption{Extrapolation performance for rectangular arrays.  Accumulation of points near vertical axis indicates the presence of persistent ambiguities with certain sampling patterns.}
\label{rect_extrap_fig}
\end{figure}

To validate performance in simulation, we tested our algorithm on small arrays with 3 rows and 3 columns of elements, spaced randomly up to one-half wavelength.  Random complex excitations were applied to each element.  A typical antenna pattern, its principal plane patterns, and extrapolation results are shown in Figure \ref{rect_eg_fig}.  Typical performance of our algorithm (using 10 iterations on 100 sample antennas) on these rectangular arrays is shown in Figure \ref{rect_extrap_fig}.  It is clear that the algorithm works well: if the residual error is low, the total error will usually be low also.  However, there is substantial dependency on the sampling pattern.  Principal plane patterns do not exhibit a definite reduction in total error, even with essentially no residual errors.  This is due to the vertical and horizontal symmetry of the array, which leads to ambiguities in the resulting pattern when sampled in the azimuthal and elevation planes.  Clearly, a lower bound on total error is achieved with random sampling.

\begin{figure}
\begin{center}
\includegraphics[width=3in]{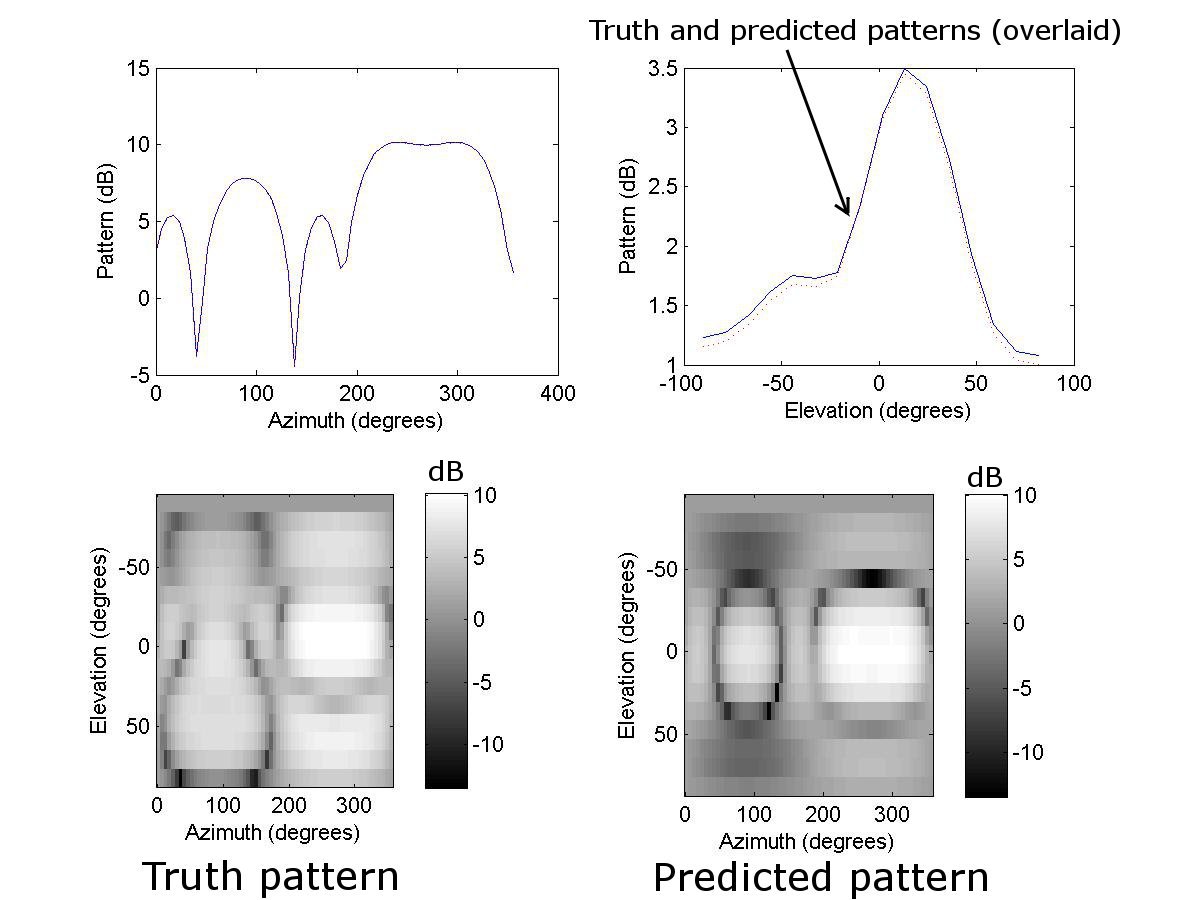}
\end{center}
\caption{Incorrectly predicted off-axis sidelobe levels for a rectangular array, due to symmetric sampling.  Top frames show azimuth and elevation patterns respectively; dotted lines represent the closest match found by our algorithm.  Bottom left frame shows truth pattern, bottom right is the prediction from our algorithm.}
\label{rect_ambiguities_fig}
\end{figure}

It therefore is important to be aware of these symmetries when selecting collections of sample points.  In particular, using strictly azimuth or elevation patterns (or both) can result in mis-estimation of non-cardinal sidelobes as shown in Figure \ref{rect_ambiguities_fig}.  These sidelobe mis-estimations are the cause of the large total errors in Figure \ref{rect_extrap_fig}.  We therefore recommend using a different sampling pattern for rectangular arrays to avoid these issues.

Rectangular arrays present an additional challenge: the number of elements may be unknown.  However, in this case, the residual errors cannot be decreased beyond a certain amount.  This provides a discrimination mechanism: if the residual errors do not decrease given a number of iterations and random initial conditions, a different number of elements in the configuration should be tried.  
To test this methodology, we simulated an antenna with 3 columns and 4 rows of elements.  We then computed the minimal residuals with our algorithm for all configurations with between 1 and 5 rows and columns.  These residuals are shown in Table \ref{rect_shape_tab}.  It should be clear that using more rows in the proposed configuration than are actually present will result in low residuals.  Therefore, the correct configuration is the one whose residuals are the smallest and has the smallest number of rows and columns.  In our table, this occurs at precisely the correct number of elments (3 columns, 4 rows).

\begin{table}
\caption{Value of smallest residual given varying numbers of rows and columns of elements (truth antenna has 3 columns and 4 rows)}
\begin{center}
\begin{tabular}{c|ccccc}
&1 column&2 columns&3 columns&4 columns&5 columns\\
\hline
1 row&730&469&283&154&145\\
2 rows&418&261&47&47&47\\
3 rows&397&225&2&0.5&0.5\\
4 rows&390&224&0.5&0.4&0.3\\
5 rows&360&216&0.4&0.3&0.2\\
\end{tabular}
\end{center}
\label{rect_shape_tab}
\end{table}

\subsection{Horn antennas}

There are several different kinds of horn antennas that are in typical usage.  For concreteness, consider an E-plane horn.  This kind of antenna has three design parameters (see Figure \ref{horn_conf_fig}): the width between parallel plates, the height of the throat of the horn, and the height of the mouth of the horn.  Therefore, $\Configuration=\mathbb{R}^3$.  In contrast to a phased array, the excitation of a horn antenna presents only one complex degree of freedom.

From these parameters, it is straightforward to compute the pattern $\Full$ of the horn.  For our simulation, we used the closed-form solutions given by (13-11b) or (13-11c) in \cite{Balanis_2005}.  For this antenna, Theorem \ref{thm:uniqueness} indicates that one needs to take at least $10$ measurements to obtain injective pattern maps $\Pattern$.  In practice, errors will be present so more measurements are generally desirable.

\begin{figure}
\begin{center}
\includegraphics[width=3in]{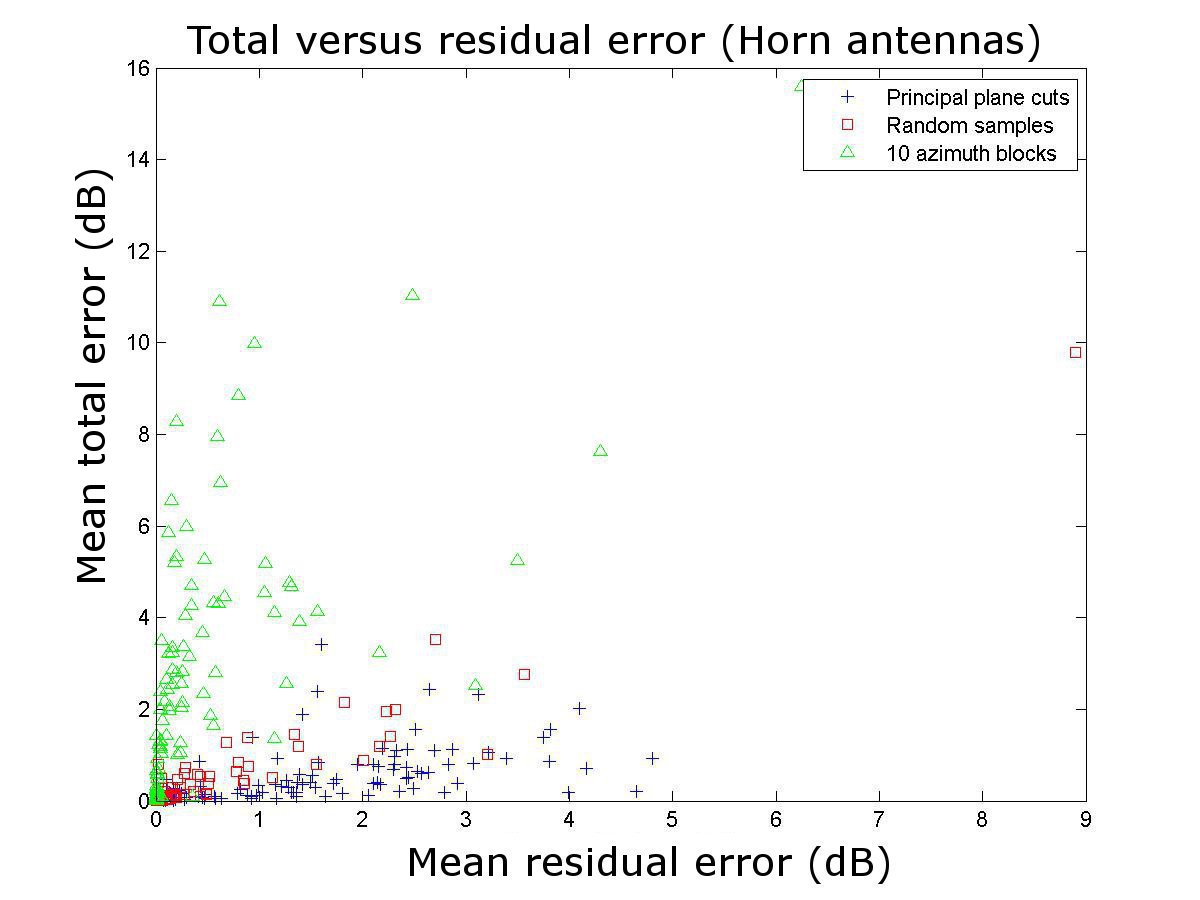}
\end{center}
\caption{Extrapolation performance for E-plane horns.  Accumulation of points near vertical axis predicted with azimuth blocks indicates the presence of an ambiguity with this sampling pattern.}
\label{horn_extrap_fig}
\end{figure}

The horns in our simulations are oriented so that the $E$-field is in the elevation plane.  Typical horns with mouth and side length dimensions up to 6 wavelengths were tested.  Overall performance of the extrapolation of E-plane horn antennas is shown in Figure \ref{horn_extrap_fig}, where 100 antennas were tested and the solver used 10 iterations at most.  Unlike the case of rectangular phased arrays, there do not appear to be any prominent symmetries that degrade performance when principal planes are used.  However, using contiguous blocks of azimuth samples (with random small elevation angles) appears result in some ambiguity.  Again, random sampling (which has the least amount of symmetry) acheives a practical lower bound on total error.

\subsection{Dish antennas}

Dish antennas can be complicated by the fact that the illumination of the dish itself is intimately connected to the feed structure.  As explained in Section \ref{sec:design}, $\Configuration=\mathbb{R}^2\times\mathbb{R}^2\times\mathbb{R}^3\times SO(3)$ (or similar) in this case, as shown in Figure \ref{dish_conf_fig}.  Unlike the case of phased arrays or horn antennas, no closed form solutions typically exist.  Therefore, the computation of $\Full$ in our algorithm will usually involve a computational electromagnetics engine.  In any case, one needs to make at least 20 measurements to obtain an injective pattern map.

To make this model concrete enough for a computer simulation, we represented the feed as an isotropic radiator.  In this case, the $SO(3)$ factor plays no role, and may be removed from $\Configuration$.  Indeed, the radiation pattern of the resulting isotropic-fed dish can be found by ray-tracing from the feed to the receiver, reflecting off each of a dense set of points on the reflector.  Specfically, the reflector is described by the equation of $y=ax^2+bz^2$.

\begin{figure}
\begin{center}
\includegraphics[width=3in]{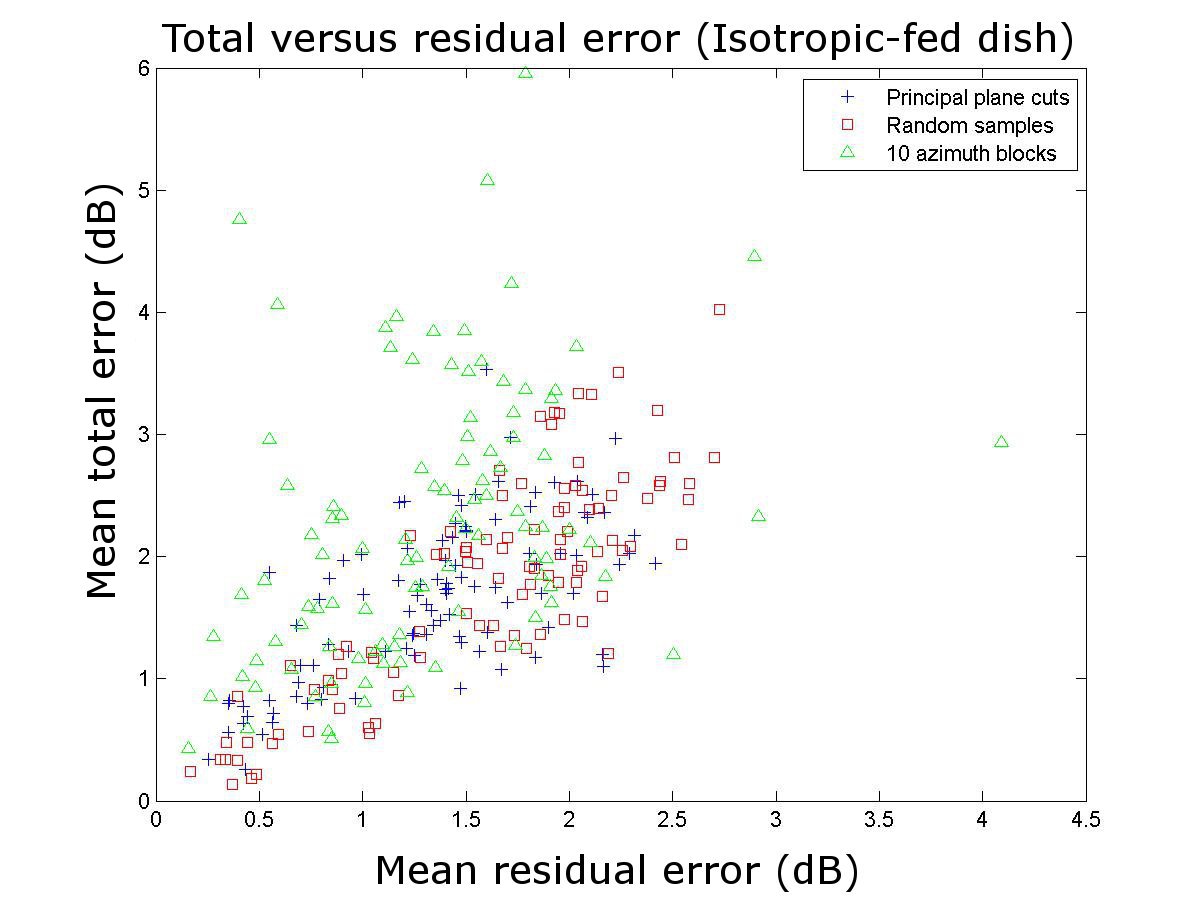}
\end{center}
\caption{Extrapolation performance for isotropic fed parabolic dishes.  No persistent ambiguities are present in the pattern predictions.}
\label{dish_extrap_fig}
\end{figure}

Our simulation tested dish antennas whose reflector radii were up to 6 wavelengths and feed was located within 6 wavelengths from the vertex of the reflector.  The parameters $a$ and $b$ were chosen randomly between 0 and 6.  The reflector was simulated with a grid with 20 axial points and 10 radial points.  The typical results for 100 dish antennas (10 iterations used in the solver) are shown in Figure \ref{dish_extrap_fig}.  Unlike both horn and rectangular arrays, the performance of dishes is essentially the same regardless of sampling pattern.  In particular, random sampling does not outperform other sampling patterns.  There are no obvious ambiguities, apparently as the model of the antenna has no symmetries that always are shared the sampling pattern.  The model can be symmetric, but pairs typical antennas will not share axes of symmetry.

\section{Conclusion}

This article presents a theoretical and algorithmic framework for extrapolating antenna radiation patterns from subsampled patterns.  Notionally, we aim to dramatically improve the speed of measurement of full patterns taken from a far-field range, allowing off-cardinal axis sidelobe structure to be estimated indirectly.  Our approach is based on the Whitney embedding theorem of differential topology, and is therefore sufficiently general to treat many different kinds of antennas with no change in algorithms.  We validated the performance of our algorithms in simulation for three different kinds of antennas: rectangular phased arrays, E-plane horns, and parabolic dishes. 



\bibliographystyle{IEEEtran}
\bibliography{antextrap_bib}

%
%
%

\begin{IEEEbiography}
[{\includegraphics[width=1in]{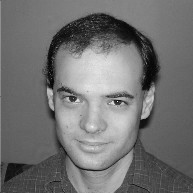}}] {Michael
  Robinson} is a postdoctoral fellow in the Department of Mathematics
at the University of Pennsylvania and a research engineer at SRC,
inc. His 2008 Ph.D. in Applied Mathematics [Cornell University] and
recent work in topological signal processing is complemented by a
background in Electrical Engineering and current work in radar systems
analysis.
\end{IEEEbiography}
\end{document}